\documentclass[aps,prb,twocolumn,floatfix,superscriptaddress]{revtex4-2}
\usepackage{graphicx}
\usepackage{amsmath,amsthm}
\usepackage{amssymb}
\usepackage{braket}
\usepackage{xcolor}
\usepackage{bm}
\usepackage{ulem}
\usepackage[bookmarks=false,linkcolor=blue,urlcolor=blue,colorlinks,citecolor=blue]{hyperref}
\usepackage{caption}
\usepackage{subcaption}

\newcommand{\be}[1]{ \begin{eqnarray} \mbox{$\label{#1}$} }

\newcommand{\ee}{\end{eqnarray}}

\newcommand{\eeq}{\end{equation}}

\newcommand\varsha[1] {{\color{black}{#1}}}

\begin{document}
\title{Geometric transport signatures of strained multi-Weyl semimetals}

\author{Varsha Subramanyan}
\email{varshas@lanl.gov}
\affiliation{Theoretical Division, T-4, Los Alamos National Laboratory, Los Alamos, New Mexico 87545, USA}

\author{Shi-Zeng Lin}
\email{szl@lanl.gov}
\affiliation{Theoretical Division, T-4, Los Alamos National Laboratory, Los Alamos, New Mexico 87545, USA}
\affiliation{Center for Nonlinear Studies (CNLS), Los Alamos National Laboratory, Los Alamos, New Mexico 87545, USA}
\affiliation{Center for Integrated Nanotechnologies (CINT), Los Alamos National Laboratory, Los Alamos, New Mexico 87545, USA}
\author{Avadh Saxena}
\email{avadh@lanl.gov}
\affiliation{Theoretical Division, T-4, Los Alamos National Laboratory, Los Alamos, New Mexico 87545, USA}
\affiliation{Center for Nonlinear Studies (CNLS), Los Alamos National Laboratory, Los Alamos, New Mexico 87545, USA}

\begin{abstract}
 The minimal coupling of strain to Dirac and Weyl semimetals, and its modeling as a pseudo-gauge field has been extensively studied, resulting in several proposed topological transport signatures. In this work, we study the effects of strain on higher winding number Weyl semimetals and show that strain is not a pseudo-gauge field for any winding number larger than one. We focus on the double-Weyl semimetal as an illustrative example to show that the application of strain splits the higher winding number Weyl nodes and produces an anisotropic Fermi surface. Specifically, the Fermi surface of the double-Weyl semimetal acquires nematic order. By extending chiral kinetic theory for such nematic fields, we determine the effective gauge fields acting on the system and show how strain induces anisotropy  and affects the geometry of the semi-classical phase space of the double-Weyl semimetal.  Further, the strain-induced deformation of the Weyl nodes results in transport signatures related to the covariant coupling of the strain tensor to the geometric tensor associated with the Weyl nodes giving rise to strain-dependent dissipative corrections to the longitudinal as well as the Hall conductance. Thus, by extension, we show that in multi-Weyl semimetals, strain produces geometric signatures rather than topological signatures. Further, we highlight that the most general way to view strain is as a symmetry-breaking field rather than a pseudo-gauge field. 
\end{abstract}
                                        
\maketitle

\section{Introduction}

Weyl semimetals are 3D topological materials that have bulk gapless points in their spectrum, the low energy Hamiltonian around which resembles a Weyl fermion. They are topologically non-trivial materials and their properties have been extensively studied in existing literature \cite{PhysRevB.83.205101,Lu_Wang_Ye_Ran_Fu_Joannopoulos_Soljačić_2015,Xu_Belopolski_Sanchez_Zhang2015,RMP, TaAs, AR,Yan_Felser_2017}. One particularly interesting aspect of Weyl semimetals is the realization of the chiral anomaly, where electron numbers with different chiralities are not \varsha{conserved} in the presence of electromagnetic fields, in condensed matter systems. This has made Weyl semimetals an exciting possible arena to obtain experimentally accessible signatures of the chiral anomaly. \cite{PhysRevB.86.115133,Jia_Xu_Hasan_2016,Ong_Liang_2021}

The application of physical strain as a way of probing and amplifying topological transport characteristics has also generated a lot of interest recently since the coupling of strain to systems with a Dirac/Weyl cone (like graphene) in the spectrum can be modelled as a pseudo-gauge potential. \cite{Vozmediano_Katsnelson_Guinea_2010,Levy_Burke_Meaker_Panlasigui_Zettl_Guinea_Neto_Crommie_2010,Cortijo,Pikulin,PhysRevX.6.041046,PhysRevB.108.125129,Chen_Zhang_Su_Cao_Xiao_Lin_2024,Su_Balatsky_Lin_2024} This property offers a way of creating large effective pseudo-electromagnetic fields through the application of strain, thus enabling us to probe such suitable materials under large field conditions. Since the strain-induced pseudo-electromagnetic fields preserve time reversal symmetry, signatures of pseudo-electromagnetic fields can be distinct from those of conventional electromagnetic (EM) fields and have also been thoroughly documented \cite{PhysRevB.96.085201,Cortijo, Pikulin,Glazman, Medel2024, GHOSH2024}. 

In Weyl semimetals, the band crossing points or Weyl nodes serve as monopole or antimonopole for the Berry curvature flux in the momentum space. Most studies to date focus on the Weyl node associated with monopole charge equal to $\pm 1$. In condensed matter systems, it is also possible to have Weyl nodes with higher monopole charge. Such higher winding number Weyl nodes have been proposed to exist in realistic materials including HgCr$_2$Se$_4$ and SrSi$_2$ for $n=2$ \cite{material1,material2,material3,SCHMELTZER2023169380}. This raises an immediate question as to how the strain couples to the Weyl fermions with higher winding number and what the consequences of this coupling are. Can strain still be treated as a pseudo-electromagnetic field in this case?

In this work, we study multi-Weyl semimetals, which are Weyl semimetals that have a winding number $n$ larger than one, and their response to strain.  We show that strain couples in fundamentally different ways to systems of higher winding number as compared to a simple Dirac or Weyl cone.  Further, we show that the behavior of strain as a pseudo-gauge field when coupling to a $n=1$ Weyl semimetal is a very specific instance of more general considerations. For example, we demonstrate that strain couples to $n=2$ Weyl semimetals as a nematic order parameter rather than a gauge field, causing a split in the higher winding number nodes to yield lower winding number nodes, in alignment with previous investigations in Ref. [\onlinecite{Sukhachov}]. Such applications of strain do not significantly alter the topology of the material, and thus do not significantly alter transport signatures that have a topological origin. Instead, they modify and couple to the geometric tensor of the system, yielding unique transport signatures not found in the simple $n=1$ semimetal. We highlight the presence of strain-dependent dissipative corrections to the longitudinal as well as the Hall conductance of multi-Weyl semimetals in this context. 

The remainder of the paper is organized as follows. We briefly introduce the notation and framework used in this work in Section \ref{S2}. We then present the nature of strain coupling and its associated field theory in Section \ref{S3}. After framing this modified strain coupling in terms of the chiral kinetic theory in Section \ref{S4}, we investigate the geometry of the system and its influence on transport signatures in Section \ref{S5}. We finally present an outlook on experimental feasibility in Section \ref{S6}.

\section{Modeling strain in multi-Weyl semimetals}\label{S2}

We begin by briefly laying out certain essential aspects of the more familiar $n=1$ Weyl semimetal and its response to strain \cite{Cortijo} in order to introduce our conventions. We consider $n=1$ Weyl semimetal with two Weyl points located at $k_z=\pm \pi/2$, which is described by the low-energy Hamiltonian
\begin{align}
    H=k_x\sigma_x+ k_y\sigma_y +(k_z\mp\frac{\pi}{2})\sigma_z,
\end{align}
with the lattice regularized Hamiltonian taking the form
\begin{align}
\begin{split}
    H=&t\sin{k_x}\sigma_x+t\sin{k_y}\sigma_y+t\cos{k_z}\sigma_z\\&+t_0(2-\cos{k_x}-\cos{k_y})\sigma_z, \label{lat1}
\end{split}
\end{align}
when the nodes are separated along $k_z$ and the energy scale is set by the hopping parameters $t$ and $t_0$. Here, the time-reversal symmetry is broken, but the inversion symmetry is preserved. Following Refs. [\onlinecite{Hughes}] and [\onlinecite{Cortijo}], strain is applied to this system as deformations in the lattice. {We consider here a generic strain represented by elements of the strain tensor $u_{ij}=\frac{1}{2}(\partial_i\delta r_j+\partial_j\delta r_i)$, where $\delta{\bf r}$ is the lattice displacement due to strain.} Effectively, this amounts to modifying the hopping terms in the following way:
\begin{align}
\begin{split}
        t_{\alpha\alpha}({\bf a_i}+\delta{\bf r_1})&=t_{\alpha\alpha}({\bf a_i})+\frac{{\bf a_i}\cdot\delta{\bf r_1}}{a_i}\frac{\partial{t_{\alpha\alpha}}}{\partial r}\Bigg|_{\bf a_i},\\
    t_{\alpha\beta}({\bf a_1}+\delta{\bf r_2})&=\frac{{\bf n}\cdot({\bf a_1}\times \delta {\bf r_2})}{a_1}t_{\alpha\beta}({\bf a_2}),
\end{split}\label{ts}
\end{align}
where $\alpha, \beta$ run over the relevant orbitals (here $s$ and $p$) and $i, j$ run over spatial directions. Here ${\bf a_i}$ are the unperturbed lattice vectors, ${\bf n}$ is a unit vector perpendicular to ${\bf a_1}$ and ${\bf a_2}$, $\delta {\bf r_1}$ is a displacement along ${\bf a_i}$ and $\delta{\bf r_2}$ is a displacement perpendicular to it. On plugging these modified hopping parameters into the lattice Hamiltonian and expanding around the Weyl node points, we obtain the modified low-energy Hamiltonian 
\begin{align}\label{eq4aa}
\begin{split}
    H&=(k_x\mp \gamma\frac{\pi}{2}u_{zx})\sigma_x+(k_y\mp \gamma\frac{\pi}{2}u_{zy})\sigma_y\\&+(k_z\mp\frac{\pi}{2} \pm  \gamma\frac{\pi}{2} u_{zz} - \gamma(u_{xx}+u_{yy}))\sigma_z, 
\end{split}
\end{align}
where {the $\pm$ sign indicates the strained Hamiltonian at each Weyl node of chirality $\pm 1$. }We have set the original hopping amplitudes to one for simplicity. The effective coupling constant $\gamma$ is the Gr\"{u}neisen parameter. That is, the effect of strain is to couple minimally to the system as an axial gauge field, or a pseudo-gauge field. Thus, strain can generate pseudo-electromagnetic fields and produce transport signatures analogous to EM fields \cite{Glazman, Pikulin}. 

We now adapt this formalism to multi-Weyl semimetals \cite{Dantas2018, GaugeAxial}. While winding numbers in a material are topological charges and thus remain topologically protected, additional point group symmetries are necessary to realize higher winding numbers within a given crystal. The necessary crystal symmetries to obtain multi-Weyl semimetals of a given winding number as well as their low-energy Hamiltonians have been discussed in Ref. [\onlinecite{Bernevig}]. Here, we focus on winding number $n=2$ systems as an illustrative case. In order to obtain a pair of $n=2$ nodes separated along the $k_z$ axis, the system must have $C_4$ or $C_6$ point-group symmetry in the $k_x$-$k_y$ plane. The low-energy Hamiltonian of such a system with $C_4$ symmetry takes the form
\begin{align}
    H&=v_1(k_x^2-k_y^2)\sigma_x+v_12k_xk_y\sigma_y+v_2k_z\sigma_z,\\
    E_0&=\pm\sqrt{v_1^2(k_x^2+k_y^2)^2+v_2^2k_z^2}.
\end{align}
Further, \varsha{ the full lattice} Hamiltonian is given  by
\begin{align}
\begin{split}
        H&=t_1(\cos{k_y}-\cos{k_x})\sigma_x+2t_1\sin{k_x}\sin{k_y}\sigma_y\\&+t_3\cos{k_z}\sigma_z+t_0(2-\cos{k_x}-\cos{k_y})\sigma_z,\label{H2}
\end{split}
\end{align}
with the Weyl nodes being located at $(0,0,\pm\frac{\pi}{2})$. The Fermi arcs - surface states associated with Weyl semimetals - now have multiple branches corresponding to the winding number \cite{Dantas2020}. 

The response of higher winding number Weyl semimetals to electromagnetic and axial fields has been studied in detail \cite{ Dantas2018,Roy, PhysRevB.96.085201} and in general, amounts to transport signatures being scaled by the winding number. However, we will now show that contrary to the previous case, strain does not couple to multi-Weyl fermions through minimal coupling as electromagnetic fields. Or in other words, the strain does not act as pseudo-electromagnetic fields for multi-Weyl semimetals.

\section{Non-minimal strain coupling in double-Weyl semimetals}\label{S3}

\begin{figure}
    \centering
    \includegraphics[width=0.5\textwidth]{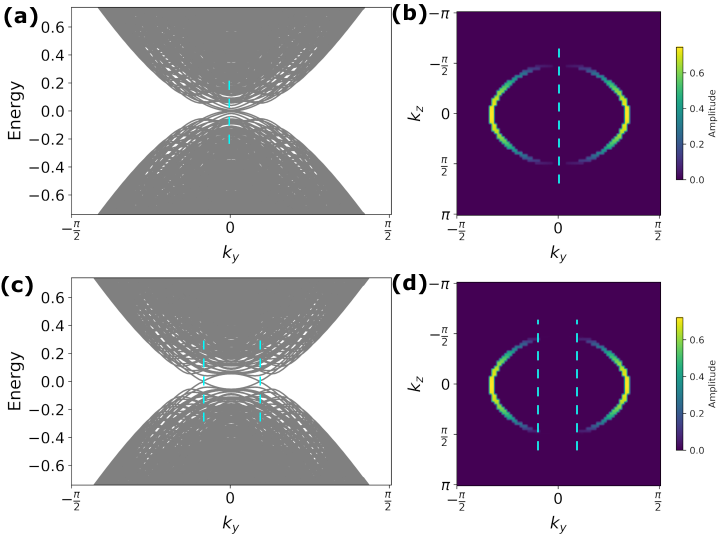}
    \caption{Band structure and surface states of the double-Weyl semimetal: This figure shows the (a) quadratic band structure of  the double-Weyl semimetal, and (b) Fermi arcs with two branches on the \varsha{left ($k_x=0$)} surface of the semimetal. The same quantities are reproduced for a strain of $u_{xx}=4\%$ along the $x$ direction showing (c) the split of each quadratic node to yield two linear nodes in the band structure, and (d) the splitting of the branches of the Fermi arc yielding two independent surface states. Dashed lines indicating the positions of the Weyl nodes in each case are also shown. \varsha{In each case, the Fermi arcs are plotted for a system of size $N=80$, discretized in the $x$ direction, as shown in the model in Eq. (\ref{disc}). The hopping strengths are all set to $t=1$. The Fermi arcs are obtained as amplitudes of the wave functions on the surface, corresponding to low-lying energies in an energy window of $\Delta E=0.05$ around zero. }}
    \label{fig:1}
\end{figure}
To describe the coupling between strain and the electrons, we start with the lattice associated with the momentum space Hamiltonian in Eq. (\ref{H2})
\begin{align}
\begin{split}
    H=\sum_{i,j,k}&\Big( -c_{i+1,j,k}^\dagger c_{i,j,k}(t_1\sigma_x+t_0\sigma_z) + t_3\sigma_z c_{i,j,k+1}^\dagger c_{i,j,k} \\&+ c_{i,j+1,k}^\dagger c_{i,j,k}(t_1\sigma_x-t_0\sigma_z)  + t_2\sigma_y c_{i+1,j+1,k}^\dagger c_{i,j,k}\\& - t_2\sigma_y c_{i+1,j-1,k}^\dagger c_{i,j,k} + 2t_0\sigma_z c_{i,j,k}^\dagger c_{i,j,k} + \textnormal{h.c.}\Big).
\end{split}
\end{align}
We again deform the lattice due to strain by modifying hopping parameters as before. By angular momentum conservation, the orbitals of interest are $s$ and $d$ for $n=2$ semimetals. The effective low-energy Hamiltonian is
\begin{align}
\begin{split}
    H&=v_1[(k_x^2-k_y^2)-\gamma(u_{xx}-u_{yy})]\sigma_x\\&+v_1(2k_xk_y-2\gamma u_{xy})\sigma_y\\&+[v_2(k_z\mp\frac{\pi}{2} \mp\gamma\frac{\pi}{2} u_{zz}) + \gamma(u_{xx}+u_{yy})]\sigma_z.\label{sth}
\end{split}
\end{align}
The $\pm$ sign indicates the strained Hamiltonian at each Weyl node of chirality $\pm 2$. \varsha{A more detailed derivation of this effective Hamiltonian is given in Appendix \ref{AppA}.} It is clear from this effective Hamiltonian that elements of the strain tensor do not couple minimally with the momentum terms in the Hamiltonian. {Further, it is notable that the strain-induced terms coupled to quadratic terms of the Hamiltonian do not change sign with chirality of the Weyl node unlike in the $n=1$ case.} That is, strain is neither a gauge field nor axial. A similar analysis can be performed for all higher $n$, showing that strain is not a pseudo-gauge field in any case where $n>1$. Therefore, a more general study of strain is needed to understand its effects on these materials. Crucially, the application of strain breaks the point group symmetries that protect the higher winding number in such materials. This is apparent from Eq. (\ref{sth}) where the modified Hamiltonian no longer has $C_4$ symmetry due to the strain-dependent terms. Therefore, the system can no longer host Weyl nodes of winding number $n=2$. Instead, each double-Weyl node splits into two simple Weyl nodes of $n=1$. It is useful here to identify the strain-dependent vector ${\bf C}=\gamma\bigg((u_{xx}-u_{yy}),\ 2 u_{xy},\ \frac{\pi}{2} u_{zz} - (u_{xx}+u_{yy})\bigg)$. Let us consider the Berry curvature around each double-Weyl node
\begin{align}
\begin{split}
        {\bf \Omega}^\pm &= \pm\Bigg(\frac{v_1^2v_2(k_x^2+k_y^2)}{2E^3}(k_x, k_y, 2k_z)\\&-\frac{v_1v_2}{2E^3}(C_xk_x+C_yk_y, C_yk_x-C_xk_y, 0)\Bigg),
\end{split}
\end{align}
where the first term corresponds to the undeformed Berry curvature and $E$ is the new band energy. The divergence of the Berry curvature thus changes from ${\bf \nabla\cdot\Omega^\pm}=\pm 2\delta({\bf k})$ to
\begin{align}
    {\bf \nabla\cdot\Omega^\pm}=\pm [\delta({\bf k-k_0})+\delta({\bf k+k_0})],
\end{align}
where ${\bf k_0}=(\rho\cos{\phi}, \rho\sin{\phi}, \pm\frac{\pi}{2}-C_z)$, $\rho^2\cos{2\phi}=C_x$ and $\rho^2\sin{2\phi}=C_y$, thus yielding four new nodes. Correspondingly, this split is also reflected in the splitting of the branches of the surface states to now have two independent Fermi arcs, as seen in Fig. \ref{fig:1}. This observation can also be easily extended to other higher winding number materials. 

It must be noted that the strain we consider here is small in comparison to the lattice vector. Therefore the distance between the split nodes is also small, and is of the order of $\sim \sqrt{|C|}$. \varsha{In reality, this number is of the order of a few percent of the lattice constant. The strained multi-Weyl semimetal thus possesses properties that are intermediate between the unstrained multi-Weyl semimetal and the simple Weyl semimetal with multiple pairs of well-separated Weyl nodes in the Brillouin zone. We investigate some of these properties in the following sections. Further, we also highlight that the strain-induced splitting of nodes does not change the net topological charge in the system, but simply redistributes it in the Brillouin zone. } 

\subsection{Director fields and effective field theory}

\begin{figure}
    \centering
    \includegraphics[width=0.5\textwidth]{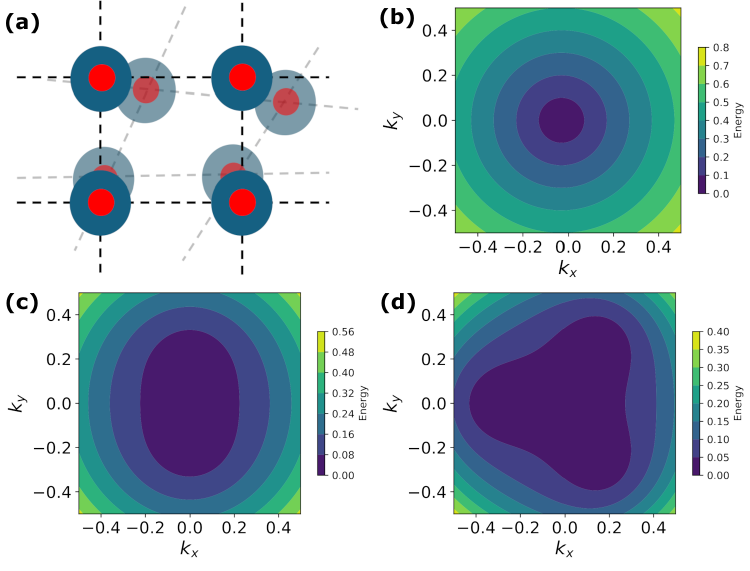}
    \caption{Contour plots of energy above the $k_z=\frac{\pi}{2}$ Weyl node, revealing the shape of the Fermi surface upon the application of strain: (a) Breaking of lattice point group symmetries by the application of strain results in the breaking of Fermi surface isotropy due to the splitting of the Weyl nodes. (b) In simple Weyl semimetals, isotropy is preserved, and strain acts as a gauge field. (c) In double-Weyl semimetals, the Fermi surface gains nematic order and strain behaves as a director field. (d) In triple- and higher winding number systems, isotropy is broken to obtain a lower $C_n$ symmetry. }
    \label{FS}
\end{figure}
Here we show that the strain can be regarded as a nematic or director field coupled to multi-Weyl fermions, and derive an effective field theory for the strain field. Without loss of generality, let us set $C_z=0$ since it only has the effect of changing the distance between the nodes. While the unstrained Hamiltonian has $C_4$ symmetry, it is easy to see that in the presence of strain, the Hamiltonian and the resultant spectrum are invariant under a $\pi$ rotation in the $k_x$-$k_y$ plane about the position of the original Weyl node, say $(0,0,\pi/2)$. The deformation of the Fermi surface in this plane is shown in Fig. \ref{FS}. 

We have already established that the vector ${\bf C}$ is not a gauge field. However, its presence that generates a $\pi$ -  rotation invariant spectrum indicates that it is a nematic order parameter field or a director field. To establish this more clearly, it is useful to consider a 2D slice of this system yielding a Dirac equation of mass $m$ with quadratic band crossing. The resultant Lagrangian is
\begin{align}
    L&=\bar{\psi}\bigg(i\partial_t\sigma_z-m\sigma_z+(\partial_x^2-\partial_y^2)\sigma_x+2\partial_x\partial_y\sigma_y+{\bf C}\cdot{\bf \sigma}\bigg)\psi.
\end{align}
This Lagrangian is identical to that of a 2D quantum anomalous Hall (QAH) system with spontaneously broken rotational symmetry leading to a nematic phase, as in Refs. [\onlinecite{Fradkin1}] and [\onlinecite{Fradkin2}] up to an overall unitary transformation. In the 2D QAH system, ${\bf C}$ is the nematic order parameter and originates due to fluctuations. In the double-Weyl semimetal system however, the role of the ``order parameter" is played by elements of the applied strain tensor that couple to the Weyl node. 

When the fermions are integrated out preserving terms upto second order in strain, we obtain the effective Lagrangian
\begin{align}
\begin{split}
        L&=-\chi \epsilon^{bc}Q_{ab}\partial_tQ_{ac}+\kappa_1\textnormal{Tr}[QKQ]+\kappa_2\textnormal{Tr}[QKQ\sigma_x],\label{Leff}
\end{split}
\end{align}
where $Q=C_x\sigma_x+C_y\sigma_y$ and the matrix $K$ is a matrix of quadratic derivatives given by 
\begin{align*}
    K&=\frac{\mathbb{I}}{2}(\partial_x^2+\partial_y^2)+\frac{\sigma_x}{2}(\partial_x^2-\partial_y^2)\\&+\frac{\sigma_y}{2}(\partial_x\partial_y+\partial_y\partial_x)+i\frac{\sigma_z}{2}(\partial_x\partial_y-\partial_y\partial_x).
\end{align*}
The corresponding Berry curvature contribution to Hall viscosity is identical to that of the 2D QAH system and is evaluated as \cite{Fradkin1} 
\begin{align}
    \chi=\frac{1}{8\pi m}.\label{qah}
\end{align}
It is useful to compare Eq. (\ref{Leff}) to the one obtained for strained $n=1$ Weyl semimetals. Since strain couples to this system as a gauge field, the effective action is simply the Chern-Simons action \cite{Volovik1,VOLOVIK2022168998,Nissinen2018,Cortijo}
\begin{align}
    L_{n=1}=-\frac{\textnormal{sgn}(m)}{2}\epsilon^{ab}A_a\partial_tA_b+ \frac{\textnormal{sgn}(m)}{2} \epsilon^{abc}A_a\partial_b A_c,
\end{align}
where ${\bf A}$ is the strain-induced pseudo-gauge field. {For the Hamiltonian in Eq. \eqref{eq4aa}, $A_j\sim \gamma u_{zj}$.} While the two effective Lagrangians share similarity in form, the effective action for the double-Weyl semimetal is not a gauge theory. Instead, it is equivalent to the Frank free energy functional for a nematic fluid with the director ${\bf C}$.

\subsection{3D Hall viscosity}
Here we evaluate the electron Hall viscosity $\eta^{3D}$ in strained Weyl semimetals. The strain-induced contribution to 3D Hall viscosity in $n=1$ semimetals is evaluated by treating the 3D Weyl semimetal as a stack of 2D topological systems layered such that they are in a weak topologically insulating phase. The mass term $m$ is replaced by $t\cos{k_z}+t_0(2-\cos{k_x}-\cos{k_y})$ and integrated through the Brillouin zone at each high symmetry point to obtain the 3D Hall viscosity term {since the 2D massive Dirac points are present at these high symmetry points. The sum of contributions at each high symmetry point for each value of $k_z$ would give us the 3D contribution to Hall viscosity}. These points in the notation of Eq. (\ref{lat1}) as well as their contributions to Hall viscosity are given below. We have set all hopping parameters to one for convenience of notation.
\begin{align*}
    (0,0)&: \frac{1}{2}\textnormal{sgn}(\cos k_z),\quad
    (0,\pi): -\frac{1}{2}\textnormal{sgn}(\cos k_z + 2),\\
    (\pi,0)&: -\frac{1}{2}\textnormal{sgn}(\cos k_z + 2),\quad
    (\pi,\pi): \frac{1}{2}\textnormal{sgn}(\cos k_z + 4).
\end{align*}
The sgn function in the latter three terms is always positive. Thus the total contribution to the effective 3D Lagrangian is 
\begin{align}
\begin{split}
    L^{3D}_\eta &= -\int_0^{2\pi} \frac{dk_z}{2\pi} \frac{1}{2}(\cos k_z -1) \epsilon^{ab}A_a\partial_tA_b\\
    &= \frac{\pi}{2\pi}\epsilon^{ab}A_a\partial_tA_b,
    \end{split}
\end{align}
where the $\pi$ in the numerator corresponds to the distance between the Weyl nodes along $k_z$. 

We can perform a similar analysis for the $n=2$ system. Analogous contributions from high symmetry points in the notation of Eq. (\ref{H2}) are given by
\begin{align*}
    (0,0)&: \frac{2}{8\pi \cos k_z},\quad
    (\pi,\pi): \frac{2}{8\pi (\cos k_z+4)}.
\end{align*}
The total contribution to the effective 3D Lagrangian is 
\begin{align}
\begin{split}
    L^{3D}_\eta &= -\int_0^{2\pi} \frac{dk_z}{2\pi}\bigg(\frac{2}{8\pi \cos k_z}+\frac{2}{8\pi (\cos k_z+4)}\bigg) \epsilon^{bc}Q_{ab}\partial_tQ_{ac}\\
    &=\frac{1}{4\pi\sqrt{15}}\epsilon^{bc}Q_{ab}\partial_tQ_{ac}.
    \end{split}
\end{align}
The first term here is nominally divergent, but one can perform a principal value analysis to set it to zero. As a direct extension, if the Weyl nodes are located at positions $k_z=\pm \theta$, then the relevant viscosity coefficient is 
\begin{align}
    \eta^{3D}&=\frac{\sqrt{2}}{4\pi\sqrt{31+\cos 2\theta-16\cos\theta}}.
\end{align}

\subsection{Field theory for higher winding numbers}
From the analysis presented in Section \ref{S2}, it is clear that elements of the strain tensor will not couple minimally to momentum like gauge fields for any winding number larger than one. Instead, they will behave as a symmetry breaking field in the lattice. This breaking of 
 real-space symmetry is further reflected in the breaking of momentum-space symmetry due to the split Weyl nodes. The Fermi surface around the split Weyl nodes is no longer isotropic for larger winding numbers. Instead, the isotropy is broken to obtain a smaller symmetry group of $n$-fold rotation or $C_n$. Contour plots of the Fermi surface obtained at a small energy just above the strained Weyl nodes are shown in Fig. \ref{FS}. 

 In the case of double-Weyl semimetals, we have demonstrated that the effective field theory obtained in the presence of strain is mapped to the theory of nematicity in quantum Hall fluids, where the Fermi surface possesses a $C_2$ symmetry. In the case of higher winding numbers, a similar effective field theory can be obtained reflecting the $C_n$ symmetry of the Fermi surface. The 2D Hall viscosity term $\chi$ as defined in Eq. (\ref{Leff}) in this case takes the general form of
 \begin{align}
     \chi=\frac{1}{4\pi\sqrt{\pi}}m^{-2(1-\frac{1}{n})}\Gamma\bigg(\frac{n+1}{n}\bigg)\Gamma\bigg(\frac{3n-2}{2n}\bigg).
 \end{align}
 The effective field theory  would no longer be nematic in the case of $n> 2$, instead reflecting the $C_n$ symmetry.

 Therefore, we see that the axial gauge theory obtained in the case of $n=1$, where isotropy of the Fermi surface around the Weyl node is preserved even under strain, is a specific realization of a larger pattern of strain coupling to a given material. {Strain, when conceived most generally, is not simply a gauge field, but a symmetry breaking field that couples to and modifies the {\it geometry} of the Fermi surface associated with the system. These changes in the geometry of the Fermi surface are then reflected in modifications to the geometry of the semiclassical phase space and the semiclassical equations of motion of the strained multi-Weyl semimetal. We demonstrate this idea explicitly in the next section. }
 
\section{Chiral Kinetic Theory}\label{S4}

{The distribution of occupied electron states $f(x,k)$ is modified from the Fermi-Dirac distribution $f_0(x,k)$ under the influence of external perturbations in the form of electromagnetic fields, pseudo-gauge fields or strain. The new electron distribution is estimated by means of the Boltzmann transport equation given by 
\begin{align}
    \partial_t f + {\bf \dot{x}}\cdot\partial_{\bf x}f +{\bf \dot{k}}\cdot\partial_{\bf k}f = I[f],
\end{align}
where $I[f]$ is a collision integral. Therefore, it is necessary to write down appropriate equations of motion to solve the Boltzmann equation and hence identify the response to a particular perturbation. In Weyl semimetals, the equations of motion in phase space are obtained from the path integral through the chiral kinetic theory \cite{NY1,CKT,Son}. In simple Weyl semimetals with $n=1$, since strain couples as a pseudo-gauge field, the associated kinetics are analogous to those produced by EM fields. In this section, we reformulate the chiral kinetic theory for multi-Weyl semimetals in order to identify the unique effects of strain in such systems. We assume that there are no electromagnetic fields present, since their effect on the equations of motion is well-known.  }

Most generally, the elements of the strain tensor can be spatially varying. Thus, we can write the effective phase space Hamiltonian (in dimensionless units) of the double-Weyl semimetal as
\begin{align}
\begin{split}
        H&=[(k_x^2-k_y^2)-C_x(x_i)]\sigma_x+[2k_xk_y-C_y(x_i)]\sigma_y+k_3\sigma_z\\&\equiv ({\bf Z}-{\bf C})\cdot{\bf \sigma}\equiv ({\bf F}\cdot {\bf \sigma}) .\label{path}
\end{split}
\end{align}
Therefore the spectrum is $E(x_i,k_i)=|F|$. The path integral can be written as
\begin{align}
    \bra{f}e^{iH(t_f-t_0)}\ket{i}&=\int Dk Dx \exp{i\int_{t_0}^{t_f}({\bf k}\cdot\dot{{\bf x}}-{\bf F}\cdot{\bf \sigma})dt}.
\end{align}
We will use the formalism laid out in, for example, Ref. [\onlinecite{CKT}] to diagonalize this Hamiltonian. Let the Hamiltonian be diagonalized by the SU(2) matrix $V(x_i,k_i)\equiv V(x,k)$ as $V^\dagger(x,k)({\bf F}\cdot{\bf \sigma}) V(x,k)=E \sigma_3$ where we have omitted the spatial index for simplicity of notation. We can insert pairs of the unitary operators into the path integral to diagonalize the Hamiltonian along the path in the following way. The subscript $t$ of any quantity denotes the value of the quantity at time instant $t$:
\begin{align*}
\begin{split}
   \dots V(k_2,x_2)&V^\dagger(k_2,x_2)e^{(-i{\bf F_2}\cdot{\bf \sigma}\Delta t)}V(k_2,x_2)V^\dagger(k_2,x_2)\\
    V(k_1,x_1)&V^\dagger(k_1,x_1)e^{(-i{\bf F_1}\cdot{\bf \sigma}\Delta t)}V(k_1,x_1)V^\dagger(k_1,x_1)\dots 
\end{split}\\
    = \dots& V(k_2,x_2)e^{(-iE_2\sigma_3\Delta t)} V^\dagger(k_2,x_2)\\&\times V(k_1,x_1)e^{(-iE_1\sigma_3\Delta t)} V^\dagger(k_1,x_1)\dots .
\end{align*}
Thus, we need to estimate the product of matrices $V^\dagger(k_2,x_2)V(k_1,x_1)$. Let $(k_1,x_1)$ and $(k_2,x_2)$ be such that $\Delta {\bf F} = {\bf F}_2-{\bf F}_1$ is small. Therefore, we can approximate
\begin{align}
    V^\dagger(k_2,x_2)V(k_1,x_1)&=e^{-i{\bf A}\cdot\Delta{\bf F}},\\
    {\bf A}&=iV^\dagger(k,x){\bf \nabla}_F V(k,x),
\end{align}
where ${\bf A}$ is SU(2) gauge connection, which emerges after SU(2) rotation of $\mathbf{F}$. Since ${\bf F}\equiv {\bf F}(k,x)$, we can expand the dot product in the exponential in the following way
\begin{align*}
   A_i\Delta F_i&=iV^\dagger\Bigg(\frac{\partial x_j}{\partial F_i}\partial_{x_j}+\frac{\partial k_j}{\partial F_i}\partial_{k_j}\Bigg)V\Bigg(\frac{\partial F_i}{\partial x_m}dx^m+\frac{\partial F_i}{\partial k_m}dk^m\Bigg)\\
   &=iV^\dagger \partial_{x_i} V dx^i +iV^\dagger \partial_{k_i} V dk^i\\
   &=A^x_idx^i+A^k_idk^i.
\end{align*}
Plugging this back into the path integral, we obtain the effective action
\begin{align}
   S_{eff}= \int_{t_0}^{t_f}({\bf k}\cdot\dot{{\bf x}}-E \sigma_3-{\bf A^x}\cdot\dot{{\bf x}}-{\bf A^k}\cdot\dot{{\bf k}}) dt.
\end{align}
{That is, we obtain an emergent gauge field description of the phase space dynamics, despite strain itself not being a gauge field. }While this action looks identical to the case of a Weyl semimetal with an electromagnetic field, the key difference in this case is that the gauge fields $A^k_i(k,x)$ and $A^x_i(k,x)$ are emergent and are functions of both position and momentum. We now explore their gauge properties.

In this context, a gauge transformation is enacted when the diagonalizing matrix is different, say $V(k,x)\rightarrow V(k,x)U(k,x)$. This results in the change in fields as 
\begin{align*}
    {\bf A^x}&\rightarrow U^\dagger {\bf A^x} U+ iU^\dagger{\bf \nabla}_xU,\\
    {\bf A^k}&\rightarrow U^\dagger {\bf A^k} U+ iU^\dagger{\bf \nabla}_kU.
\end{align*}
Let us assume the gauge fields of interest are Abelian (although one can repeat this analysis more carefully for non-Abelian gauge fields). Therefore for $U\sim e^{i\theta(k,x)}$, the fields transform as 
\begin{align*}
   A^x_i&\rightarrow A^x_i(k,x)+\partial_{x_i}\theta(k,x),\\
   A^k_i&\rightarrow A^k_i(k,x)+\partial_{k_i}\theta(k,x).
\end{align*}
Similarly, the derivatives of the fields transform as
\begin{align*}
    \partial_{x_i}A^x_j&\rightarrow \partial_{x_i}A^x_j -i\partial_{x_i}\theta \partial_{x_j}\theta + iU^\dagger \partial_{x_i}\partial_{x_j}U,\\
    \partial_{k_i}A^k_j&\rightarrow \partial_{k_i}A^k_j -i\partial_{k_i}\theta \partial_{k_j}\theta + iU^\dagger \partial_{k_i}\partial_{k_j}U.
\end{align*}
It is clear from these expressions that the quantities $\partial_{x_i}A^x_j-\partial_{x_j}A^x_i$ and $\partial_{k_i}A^k_j-\partial_{k_j}A^k_i$ are gauge invariant as expected since they are simply the curls of the fields: 
\begin{align}
   {\bf \nabla}_x\times {\bf A^x}&={\bf b},\label{bfield}\\
   {\bf \nabla}_k\times {\bf A^k}&={\bf \Omega}.
\end{align}

However, since these fields are functions of both $k$ and $x$, we can form yet another gauge invariant quantity. Consider the cross derivatives
\begin{align*}
    \partial_{k_j}A^x_i&\rightarrow \partial_{k_j}A^x_i-i\partial_{k_j}\theta\partial_{x_i}\theta +iU^\dagger\partial_{k_j}\partial_{x_i}U,\\
    \partial_{x_i}A^k_j&\rightarrow \partial_{x_i}A^k_j-i\partial_{x_i}\theta\partial_{k_j}\theta +iU^\dagger\partial_{x_i}\partial_{k_j}U.
\end{align*}
It is clear that the difference of these two quantities is also gauge invariant. That is, we define the third gauge invariant quantity
\begin{align}
    S_{ij}=\partial_{k_j}A^x_i-\partial_{x_i}A^k_j.
\end{align}
The elements of this tensor are cross-terms of the generalized Berry curvature. 

Along with Eq. (\ref{bfield}), we can also define $e_i=-\partial_{x_i}E$ to form analogous effective EM fields that, most generally, vary across phase space. For completeness, we specify the explicit expressions for these fields here assuming no strain-induced displacement in the $\hat{z}$-direction [See Eq. (\ref{path})]:
\begin{align*}
    A^x&=\Bigg(\frac{Z_x \partial_xC_y-Z_y \partial_xC_x}{Z(Z+Z_z)},\frac{Z_x \partial_yC_y-Z_y \partial_yC_x}{Z(Z+Z_z)},0\Bigg),\\
    A^k&=2\Bigg(\frac{-Z_xk_y+Z_yk_x}{Z(Z+Z_z)},\frac{-Z_xk_x-Z_yk_y}{Z(Z+Z_z)},0\Bigg),\\
    e&=\Bigg(\frac{Z_x}{Z} \partial_xC_x+\frac{Z_y}{Z} \partial_xC_y,\frac{Z_x}{Z} \partial_yC_x+\frac{Z_y}{Z} \partial_yC_y,\frac{Z_z}{Z} \partial_zC_z\Bigg),\\
    b&=\Bigg(0,0,\frac{Z_z}{Z}( \partial_xC_x \partial_yC_y- \partial_xC_y \partial_yC_x)\Bigg).
\end{align*}
\varsha{Therefore, there exist strain-induced emergent electromagnetic fields in multi-Weyl semimetals as well, analogous to the pseudo-gauge potentials in the $n=1$ case. However, unlike the $n=1$ case, strain is not minimally coupled to momenta and hence, the gauge potentials are not directly proportional to strain. Instead, they are obtained through the semi-classical theory outlined in this section. Here, the gauge fields generally have both momentum and spatial dependence. This analysis can be trivially extended to the higher winding number Weyl materials as well. We now obtain the equations of motion induced by these emergent fields.  }

 \subsection{Equations of motion}
As usual, defining $v_i=\frac{\partial E}{\partial k_i}$, we write down the equations of motion from the Lagrangian to obtain the following
\begin{align}
    \dot{k}_j(\delta_{ij}+S_{ij})&= e_i-(\dot{{\bf x}}\times b)_i,\\
    \dot{x}_j(\delta_{ij}+S^T_{ij})&= v_i+(\dot{{\bf k}}\times \Omega)_i.
\end{align}

These equations of motion are exactly analogous to those obtained in the case of the anomalous Hall effect, as well as the simple Weyl semimetal except for the presence of terms involving the tensor $S_{ij}$. These terms are indicative of the anisotropic nature of the phase space in the presence of strain. Further, it is useful to compare these equations of motion to the generalized form $\omega_{\alpha\beta}\dot{\xi}^\beta=\partial_\alpha H$ where the 2-form $\omega=\frac{1}{2}\omega_{\alpha\beta}d\xi^\alpha\wedge d\xi^\beta$ is the symplectic form of this manifold \cite{Faddeev}. We read off its elements to yield
\begin{align}
    \omega_{\alpha\beta}=
        \begin{pmatrix}
        \epsilon_{ijk}b_k&-\delta_{ij}-S_{ij}\\
        \delta_{ij}+S_{ij}^T&\epsilon_{ijk}\Omega_k
        \end{pmatrix},
\end{align}
where the indices $i,j$ run from 1 to 3, making this a 6-dimensional matrix. {The symplectic form $\omega=\frac{1}{2}\omega_{\alpha\beta}d\xi^\alpha\wedge d\xi^\beta$ fully determines the geometry of the phase space manifold of the strained double-Weyl semimetal, and its dependence on strain demonstrates the ways in which this geometry is modified due to strain. It is particularly useful to highlight the strain-induced modifications to the phase space density as well as Poisson brackets defined over the manifold. The determinant of $\omega_{\alpha\beta}$ determines the phase space density $\Phi$ of the manifold \cite{Faddeev,Fujikawa} given by
\begin{align}
    \Phi=\sqrt{\det(\omega_{\alpha\beta})}=1-\Omega_ib_i+S_{ii}.
\end{align}
The term $S_{ii}$ is a function of both position and momentum and highlights the strain-induced anisotropy of phase space volume in multi-Weyl semimetals. There is no analogous anisotropy for simple Weyl semimetals.} Further, the symplectic form is also closely related to the Poisson bracket defined on the manifold. Explicitly written out, the Poisson bracket over such a manifold is written as $\{f,g\}=\omega^{\alpha\beta}\partial_\alpha f\partial_\beta g$ and thus depends on the inverse of the symplectic form. Preserving only the terms quadratic or linear in strain, we can estimate the most common Poisson bracket in the following manner
\begin{align}
    \{x^i,k_j\}&=\frac{\delta^i_{j}-b^i\Omega_j}{(1-\Omega\cdot b)}-\frac{S_{ij}^T}{(1-\Omega\cdot b)^2}.
\end{align}
{Thus, we have shown the semiclassical manifestations of the distortions induced by strain on the geometry of the phase space. Any analysis of transport must therefore take into account these distortions. However, the geometry of the underlying semimetal manifests itself in transport signatures more transparently through the quantum geometric tensor. This is highlighted in the following section.}

\section{Transport signatures and the quantum geometric tensor}\label{S5}

\begin{figure*}
     \centering
     \begin{subfigure}[b]{0.32\textwidth}
         \centering
         \includegraphics[width=\textwidth]{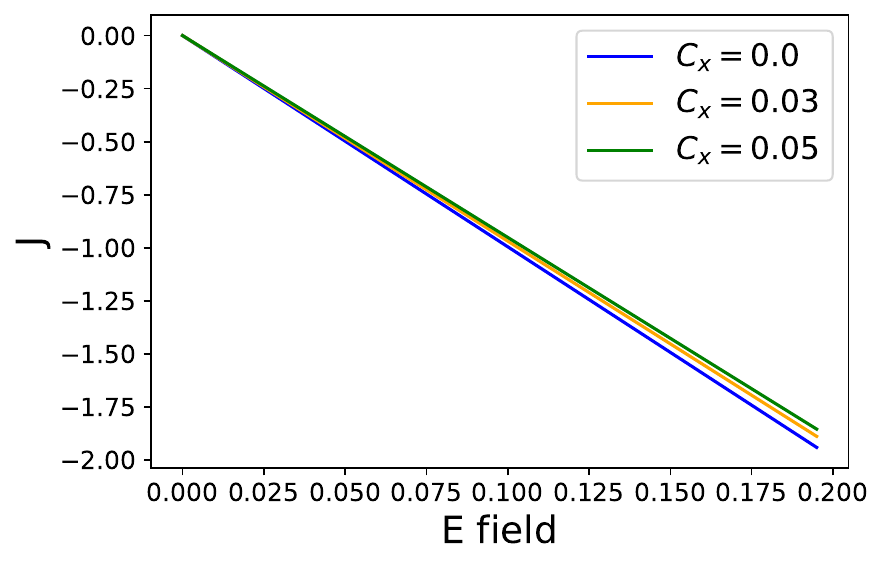}
         \caption{}
         \label{fig:y equals x}
     \end{subfigure}
     \hfill
     \begin{subfigure}[b]{0.32\textwidth}
         \centering
         \includegraphics[width=\textwidth]{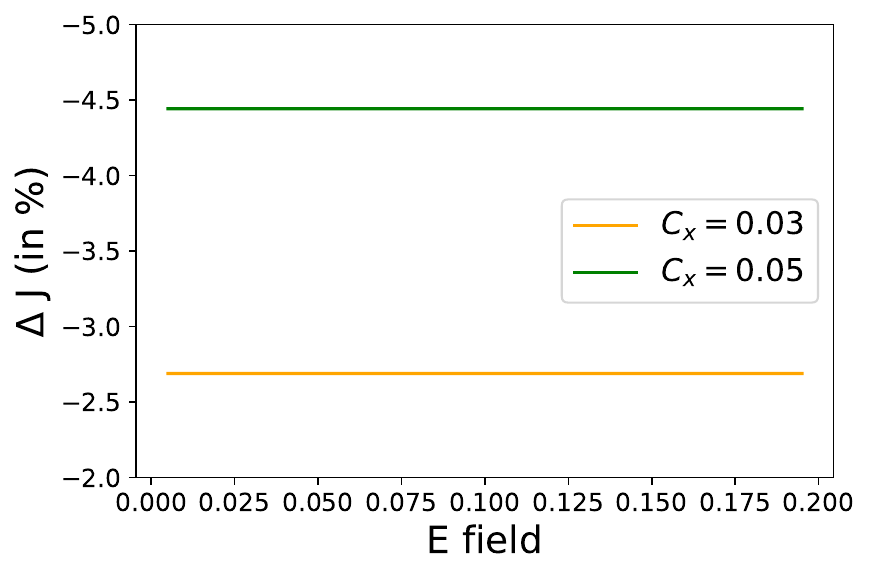}
         \caption{}
         \label{fig:three sin x}
     \end{subfigure}
     \hfill
     \begin{subfigure}[b]{0.32\textwidth}
         \centering
         \includegraphics[width=\textwidth]{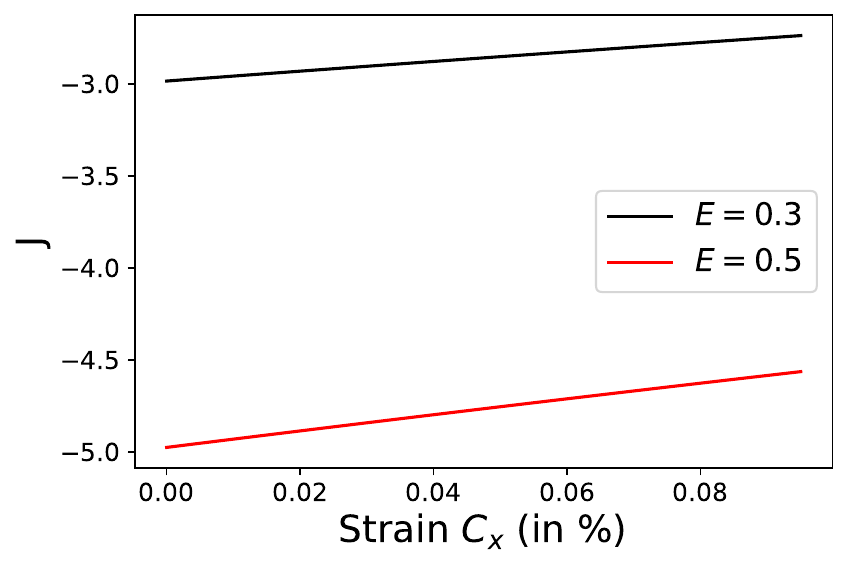}
         \caption{}
         \label{fig:five over x}
     \end{subfigure}
        \caption{Corrections to current transport in double-Weyl semimetals due to a constant strain: (a) Current (in units of $et/\hbar$) is shown as a function of electric field (in units of $t/e$) in the presence of applied strain (expressed as a fraction of hopping parameter $t$). Applied strain modifies the current in the system. (b) Difference between current in the strained system and the unstrained system are plotted as a function of electric field. Applied strain thus modifies current by a constant fraction of the current in the unstrained semimetal. (c) Current is plotted at a constant electric field for varying strengths of applied strain. Current thus varies linearly with strain for small values of strain.}
        \label{Current}
\end{figure*}

As we have discussed in earlier sections, the splitting of the  Weyl nodes due to strain does not significantly alter the topology of the material, but instead induces anisotropy in the geometry of the phase space and Fermi surface. Therefore, contributions to transport (induced by external EM fields) that are topological in origin are not significantly altered. For instance, consider the Hall current in the system, which is given by \cite{CKT, Son, Dantas2018}
\begin{align}
    J_{Hall}&=e^2 {\bf \tilde{E}}\times\int_{\bf k} f\Omega,
\end{align}
where $f$ is the modified electron distribution and ${\bf\tilde{E}}$ is the electric field. The 3D Hall conductance thus obtained is proportional to the distance between the Weyl nodes. This quantity is only modified by the small strain component $C_z$ and thus is not significantly different from its unstrained counterpart. A similar argument can be made regarding the current contributions due to the chiral anomaly as seen in, for instance, the chiral magnetic effect. While strain alters the density of states that determines the current, and thus leads to corrections to the current, these contributions are unsurprising effects of the anisotropy that has been introduced. 

Therefore, one must look at transport signatures that originate due to the modified geometry of the system, which does not exist in the unstrained material. For simplicity, let us consider a constant strain, and examine the change in energy of the bands and its effects on transport. Unless otherwise specified, we only retain terms linear in strain.
\begin{align*}
   & E=\sqrt{[v_1(k_x^2-k_y^2)-C_x]^2+(2v_1k_xk_y-C_y)^2+v_2^2k_z^2}
   \\& \sim \sqrt{v_1^2(k_x^2+k_y^2)^2+v_2^2k_z^2-2C_xv_1(k_x^2-k_y^2)-2C_yv_1(2k_xk_y)}\\
    &\sim E_0 - {\bf C}.\frac{{\bf Z}}{E_0}
    \equiv E_0+\delta E,
\end{align*}
where $E_0=\sqrt{v_1^2(k_x^2+k_y^2)^2+v_2^2k_z^2}$ and $Z_i$ are the elements of the Hamiltonian as defined in Eq. (\ref{path}). We have set $C_z=0$ again. The conduction current associated with this strain would be given by
\begin{align}
\begin{split}
        J_i&=e\int \frac{d^3 k}{(2\pi)^3} v_i(E) f(E-\mu)\\&=e\int \frac{d^3 k}{(2\pi)^3} (v_i(E_0) + \delta v_i)f(E_0+\delta E-\mu),\label{corr}
\end{split}
\end{align}
where $v_i(E_0)=\partial_{k_i}E_0/\hbar$ and $\delta v_i=\partial_{k_i}\delta E/\hbar$. That is, all additional contributions to conduction current are functions of $\delta E$. This quantity, however, is implicitly dependent on the geometry around the Weyl nodes, which is seen when $\delta E$ is written in terms of the geometric tensor associated with the system. 

The deformation of the Weyl cone points to a geometric description of the strain coupled to Weyl fermions. For this purpose, we introduce quantum geometry tensor, which is defined as \cite{provost1980riemannian}
\begin{equation}
\eta_{ab}(\mathbf{k}) = {\langle \partial_a u_\mathbf{k}| \partial_b u_\mathbf{k}\rangle}-{\langle \partial_a u_\mathbf{k}|u_\mathbf{k}\rangle\langle u_\mathbf{k}| \partial_b u_\mathbf{k}\rangle},
\end{equation}
where $u_\mathbf{k}(\mathbf{r})$ is the periodic part of the Bloch function, $\partial_a \equiv \partial_{k_a}$. Its real part is the quantum metric, $g_{ab}(\mathbf{k}) = \Re\left(\eta_{ab}(\mathbf{k})\right)$ which characterizes the geometric properties of the electron wave functions in the momentum space. The imaginary part is the more familiar Berry curvature, $\Omega_{ab}(\mathbf{k}) = -2\Im\left(\eta_{ab}(\mathbf{k})\right)$, and is responsible for the topological properties of electronic systems. While the role of Berry curvature has been well appreciated, recently, it has been recognized that the quantum metric of the electron wave function plays an increasingly important role in determining the physical properties. \cite{PhysRevLett.127.277201,PhysRevLett.112.166601,PhysRevLett.132.026301,Gao_Liu_Qiu_Ghosh2023,PhysRevLett.122.227402,PhysRevB.99.121111,peotta2015superfluidity,torma2022superconductivity,parameswaran2013fractional,Bergholtz_Liu_2013,neupert2015fractional,LIU2024515,PhysRevResearch.5.L032022,PhysRevLett.132.096602,PhysRevB.110.035130,shavit2024quantum,Jahin_Lin_2024,PhysRevResearch.6.L032063} Below we show that the strain couples to the quantum metric of the double-Weyl semimetal.

The elements of the geometric tensor associated with the double-Weyl semimetal can be obtained by using the definition, and are given by:
\begin{align}
\begin{split}
    g_{xx}&= 4v_1^2\frac{(k_x^2+k_y^2)}{E_0^4}[v_1^2(k_x^2+k_y^2)k_y^2+v_2^2k_z^2],\\
    g_{yy}&= 4v_1^2\frac{(k_x^2+k_y^2)}{E_0^4}[v_1^2(k_x^2+k_y^2)k_x^2+v_2^2k_z^2],\\
    g_{xy}&= -4v_1^4\frac{k_xk_y(k_x^2+k_y^2)^{2}}{E_0^4}.
\end{split}
\end{align}
We will ignore the effects of strain on the geometric tensor itself since it would lead to higher order corrections in the quantities that follow. In terms of the geometric tensor, we can write the change in energy as
\begin{align}
\begin{split}
    \delta &E=-v_1\bigg(\frac{C_x}{E_0}\frac{(g_{xx}-g_{yy})}{W_2^2}+\frac{C_y}{E_0}\frac{2g_{xy}}{W_2^2}\bigg)\\
    &=-v_1\bigg(\frac{(u_{xx}-u_{yy})}{E_0}\frac{(g_{xx}-g_{yy})}{W_2^2}+\frac{2u_{xy}}{E_0}\frac{2g_{xy}}{W_2^2}\bigg),
    \end{split}
\end{align}
where we have defined $W_2=2v_1^2\frac{(k_x^2+k_y^2)}{E_0^2}.$ That is, the change in energy around each Weyl node is a function of the covariant coupling between the strain tensor and the geometric tensor. Thus, as seen in Eq. (\ref{corr}), all strain-induced corrections to the current are functions of this coupled geometric term. 

A similar expression for $\delta E$ as a function of elements of the geometric tensor is obtained here for higher winding number $n$. The energy dispersion for such a semimetal is given by
\begin{align}
    E_0&=\pm\sqrt{v_1^2(k_x^2+k_y^2)^n+v_2^2k_z^2}.
\end{align}The relevant terms of the geometric tensor can themselves be easily generalized in the following way
\begin{align}
\begin{split}
    g_{xx}&\sim n^2v_1^2\frac{(k_x^2+k_y^2)^{n-1}}{E_0^4}[v_1^2(k_x^2+k_y^2)^{n-1}k_y^2+v_2^2k_z^2],\\
    g_{yy}&\sim n^2v_1^2\frac{(k_x^2+k_y^2)^{n-1}}{E_0^4}[v_1^2(k_x^2+k_y^2)^{n-1}k_x^2+v_2^2k_z^2],\\
    g_{xy}&\sim -n^2v_1^4\frac{k_xk_y(k_x^2+k_y^2)^{2(n-1)}}{E_0^4}.
\end{split}
\end{align}
The change in energy for all $n$ is given by
\begin{align}
\begin{split}
    \delta E=-v_1\bigg(\frac{C_x(u_{ij},n)}{E_0}&\textnormal{Re}(k_x+ik_y)^n\\&+\frac{C_y(u_{ij},n)}{E_0}\textnormal{Im}(k_x+ik_y)^n\bigg),
\end{split}
\end{align}
where $C_x(u_{ij},n)$ and $C_y(u_{ij},n)$ are combinations of terms from the strain tensor that couple appropriately for a particular $n$. Since the geometric tensor is always even under inversion, there is a difference between the nature of coupling for even and odd $n$. On examining the structure of $(k_x+ik_y)^n$ for even $n=2p$, we have 
\begin{align*}
    (k_x+ik_y)^n&=(k_x+ik_y)^{2p}=(k_x^2-k_y^2+i2k_xk_y)^p\\
    &=\bigg(\frac{-n^2}{4W_n^2}\bigg)^p(g_{xx}-g_{yy}+i2g_{xy})^p,
\end{align*}
where $W_n=\frac{n^2}{2}\frac{v_1^2(k_x^2+k_y^2)^{n-1}}{E_0^2}$. Therefore, for all even $n$, we can represent the change in energy, and thus the transport current in terms of the geometric tensor. Performing a similar analysis for odd $n=2p+1$, we have
\begin{align*}
    (&k_x+ik_y)^{2p+1}=(k_x^2-k_y^2+i2k_xk_y)^p(k_x+ik_y)\\
    &=\bigg(\frac{-n^2}{4W_n^2}\bigg)^p(g_{xx}-g_{yy}+i2g_{xy})^p\bigg(\frac{n}{2E_0}\frac{v_x}{W_n}+i\frac{n}{2E_0}\frac{v_y}{W_n}\bigg),
\end{align*}
where the change in energy will depend on not just the geometric tensor but also the velocity. It is interesting to note here that there is no coupling to the geometric tensor in the case of $n=1$. In summary, for all $n>1$, the change in energy $\delta E$ is always a function of the geometric tensor and the strain tensor.

\subsection{Modifications to conductivity}

In the presence of an electric field, one can obtain further corrections to conduction current linear in strain and dependent on $\delta E$. We apply an electric field ${\bf \tilde{E}}$ to the strained semimetal. Using the relaxation time approximation, we have
\begin{align}
    f(E)=f_0(E)-e\tau {\bf \tilde{E}}.({\bf v}+\delta {\bf v})\partial_Ef_0(E),
\end{align}
where $\tau$, the effective scattering time, is treated here simply as a parameter. Up to leading order in strain, the response current is evaluated in the following way
\begin{align}
\begin{split}
    J_i&=-e^2\tau \int \frac{d^3 k}{(2\pi)^3} (v_i+\delta v_i){\bf \tilde{E}}.({\bf v}+\delta{\bf v})\partial_Ef_0\\
    &\sim -e^2\tau \int \frac{d^3 k}{(2\pi)^3} v_i {\bf \tilde{E}}.{\bf v} \partial_Ef_0 -e^2\tau \int \frac{d^3 k}{(2\pi)^3} \delta v_i {\bf \tilde{E}}.{\bf v} \partial_Ef_0 \\&-e^2\tau \int \frac{d^3 k}{(2\pi)^3} v_i {\bf \tilde{E}}.\delta {\bf v} \partial_Ef_0.
\end{split}
\end{align}
Evaluating this expression explicitly with restored units yields corrections to $\sigma_{xx}$ as well as a non-zero dissipative Hall conductance. 
\begin{align}
\begin{split}
    J_x
    \sim -\frac{e^2\tau}{2\hbar^2v_2} \tilde{E}_x\bigg(&\frac{2}{3\pi^2}\mu^2 -\frac{C_x}{2\pi}\mu\bigg)\\&+\frac{e^2\tau}{2\hbar^2v_2} \tilde{E}_y\bigg(\frac{C_y}{4\pi^2}\mu -\frac{3C_y}{16\pi}\mu\bigg).\label{Corr}
\end{split}
\end{align}
A similar expression holds for $J_y$. Note that this is not the topological 3D Hall conductance which is not dissipative, and requires breaking time reversal. Here, we see that $C_x=\gamma(u_{xx}-u_{yy})$ strain brings about corrections to regular conductance and $C_y=2\gamma u_{xy}$ brings out corrections to Hall conductance. 

By employing a lattice version of the $n=2$ semimetal, we verify the linear order current produced in this system. {Here, we use real space description in the $x$ direction (denoted by the site index $j$), and use momentum space description in the $k_y$ and $k_z$ directions.} The Hamiltonian of interest is thus
\begin{align}
\begin{split}
    H=\sum_{j,k_y,k_z}\bigg( &-c_{j+1}^\dagger c_{j}(t\sigma_x+t\sigma_z +i2t\sin{k_y})\\& + c_{j}^\dagger c_{j}(t(2-\cos{k_y}+\cos{k_z})\sigma_z) + \textnormal{h.c.}\bigg).\label{disc}
\end{split}
\end{align}
It is clear from this Hamiltonian that only current along the $x$-direction is defined. Similarly, the only strain component that is relevant here is $u_{xx}$. In the presence of a small electric field, current can be estimated in units of $et/\hbar$ by direct diagonalization as
\begin{align}
    J_x=e\bigg\langle\frac{d n_j}{dt}\bigg\rangle=\frac{e}{i\hbar N}\sum_{j,k_y,k_z}[c_j^\dagger c_j,H],
\end{align}
where $N$ is the length of the lattice. The current thus obtained is shown in Fig. \ref{Current}(a) for strain coupling $C_x=\gamma u_{xx}$ represented as percentages of hopping parameters, where they have all been set to $t=1$. We see that current varies linearly with electric field as expected. Further, we see that the application of strain adds corrections to the current. This correction is estimated as a percentage of the current in the unstrained semimetal in Fig. \ref{Current}(b) which is always a near constant percentage. This result is expected when we consider Eq. (\ref{Corr}) where
\begin{align}
    \frac{\Delta J_x}{J_x}&= -\frac{C_x}{2\pi}\mu \bigg/\frac{2}{3\pi^2}\mu^2=-\frac{3}{4\pi\mu}C_x,\label{frac}
\end{align}
which depends only on the energy scale set by the strain coupling \varsha{$C_x=2\gamma t_1  u_{xx}$ (See Appendix \ref{AppA})}. The linear variation of total current for small applied strains and small electric fields is also shown in Fig. \ref{Current}(c).

\section{Experimental feasibility}\label{S6}

While some synthetic materials have been proposed to realize the double-Weyl semimetal discussed in this work, the chief candidate material reported in the literature is HgCr$_2$Se$_4$. There have been several experimental and computational studies demonstrating its quadratic band structure and Chern number $n=2$ as seen through the quantum anomalous Hall effect \cite{material1,material2,Singh2018,m4,m5,m3,material3}. A full eight band $k.p$ analysis of its band structure is given in Ref. [\onlinecite{material1}]. It is useful to focus on the low-energy effective Hamiltonian of the form
\begin{align}
    H=
    \begin{pmatrix}
        M&Dk_zk_+^2\\Dk_zk_-^2&-M
    \end{pmatrix},
\end{align}
where $M=M_0-\beta(k_x^2+k_y^2+k_z^2)$ and $k_\pm=k_x\pm ik_y$. The presence of $k_\pm^2$ terms is indicative of winding number $n=2$. The resultant spectrum is
\begin{align}
    E=\sqrt{M^2+D^2k_z^2(k_x^2+k_y^2)^2},
\end{align}
with Weyl nodes at $k_z=\pm\sqrt{M_0/\beta}$. The relevant coefficients as mapped on to the model used in the rest of our work are
\begin{align}
    v_1=\sqrt{\beta^2+\frac{D^2M_0}{\beta}},\quad v_2=\frac{DM_0}{\beta}.
\end{align}
From the Hamiltonian, it is clear that this model is not identical to the generic $n=2$ model used in this work. This is, however, immaterial to the key results discussed here. The breaking of point group symmetries by strain will indeed break the quadratic Weyl nodes to linear ones and distort the isotropic geometry of the nodes. The coupling of strain to the geometric tensor leading to corrections in the transport quantities is also independent of the specifics of the model and is instead a result of the quadratic band structure alone. Thus the measurements and transport signatures proposed in this work are realizable in principle in  HgCr$_2$Se$_4$. The magnitudes of the expected responses will crucially depend on the parameter $\gamma$ signifying the strength of coupling between the Weyl node and strain, as well as $v_1$ and $v_2$ which describe the low energy band dispersion.

\varsha{
It is notable here that the Gr\"{u}neisen parameter, that quantifies the coupling between strain and the lattice, determines the strength of all strain-induced transport signatures. This is reflected in our toy model of conduction current through Eq. (\ref{frac}). This parameter is dimensionless and generally reflects the change in the characteristic phonon frequencies of the solid with change in volume, and can be roughly determined based on estimates of the nature of interatomic potential of the lattice itself \cite{gp}. A true estimate of the magnitude of the transport signatures would thus require an estimate of the Gr\"{u}neisen parameter of the candidate materials. To estimate Eq. (\ref{frac}), we use the typical value of $\gamma=19/6$ for the Lennard-Jones interatomic potential in 3D \cite{Krivtsov2011}. Assuming a strain of $4\%$ and a doping such that the chemical potential $\mu$ and hopping strength $t_1$ are related by $\mu\sim 0.1 t_1$, we obtain a change in current of $\Delta J \sim 60\%$. At lower temperatures, this effect will be further enhanced by the chemical potential in the denominator.} 

\section{Discussion}
The topology of multi-Weyl semimetals as well as their transport signatures associated with topology are direct generalizations of simple Weyl semimetals. It has been shown in previous works that such signatures are simply scaled by the appropriate winding number, as we might expect. The geometric features of multi-Weyl semimetals, however, are distinct from their $n=1$ counterparts. Strain serves as a way to probe these distinct geometric features, while leaving their topology largely invariant.

In this work, the double-Weyl semimetal serves as an illustrative case study to understand the more general ways in which strain affects topological semimetals. While the material realization of the simple Weyl semimetal is not predicated on additional lattice rotational symmetries, the same is not true in the realization of multi-Weyl semimetals. The application of strain breaks these symmetries in multi-Weyl semimetals, resulting in the splitting of the Weyl nodes. Further, the rotational symmetry or isotropy in the Fermi surface is also broken, with the strained system only possessing a subgroup symmetry of  $C_n$. That is, the breaking of symmetries in the lattice is reflected in the breaking of isotropy of the Fermi surface, with strain playing a mediating role. In the absence of any necessary lattice symmetries in the case of $n=1$ semimetals, these effects of the strain simply reduce to that of the well-understood pseudo-gauge field. This picture of strain coupling as a symmetry breaking term is thus a more general way to understand strain.

Similarly, the application of strain changes the energy dispersion of each Weyl node by an amount determined by the covariant coupling of the strain tensor and the geometric tensor. This change in energy is directly associated with several transport signatures that we have evaluated here. In particular,  we have shown that in the presence of an electric field, strain modifies the conductance tensor of the system by inducing a dissipative correction in both regular conductivity as well as Hall conductivity. Thus, we have highlighted the ways in which the interaction between real-space geometry and quantum geometry influence transport in topological materials. 

\section*{Acknowledgments}
The work was carried out under the auspices of the U.S. DOE NNSA under contract No. 89233218CNA000001 through the LDRD Program, and was supported by the Center for Nonlinear Studies at LANL (LYY), and was performed, in part, at the Center for Integrated Nanotechnologies, an Office of Science User Facility operated for the U.S. DOE Office of Science, under user proposals $\#2018BU0010$ and $\#2018BU0083$.

\appendix
\section{Derivation of the strained Hamiltonian} \label{AppA}
In this appendix, we provide a detailed description of the way strain modifies the effective low-lying Hamiltonian of the double-Weyl semimetal. A similar analysis can be carried out for all higher winding number materials. Here, we roughly follow the method in Ref. [\onlinecite{Cortijo}] and [\onlinecite{Hughes}]. We begin with the unstrained full lattice Hamiltonian
\begin{align}
\begin{split}
        H&=t_1(\cos{k_y}-\cos{k_x})\sigma_x+2t_1\sin{k_x}\sin{k_y}\sigma_y\\&+t_3\cos{k_z}\sigma_z+t_0(2-\cos{k_x}-\cos{k_y})\sigma_z.
\end{split}
\end{align}
Here, the constant $2$ in the last term has been chosen to ensure lattice regularization, and that the Hamiltonian describes a material in the Weyl semimetal phase whose Weyl points are located at $(0,0,\pm\pi/2)$. For all calculations, every hopping parameter has been set to one in the rest of this work. Here, we label them differently to distinguish between them in the discrete spatial Hamiltonian given below:
\begin{align}
\begin{split}
    H=&\sum_{i,j,k}\Big( -c_{i+1,j,k}^\dagger c_{i,j,k}(t_1\sigma_x+t_0\sigma_z) + t_3\sigma_z c_{i,j,k+1}^\dagger c_{i,j,k} \\&+ c_{i,j+1,k}^\dagger c_{i,j,k}(t_1\sigma_x-t_0\sigma_z)  + t_2\sigma_y c_{i+1,j+1,k}^\dagger c_{i,j,k}\\& - t_2\sigma_y c_{i+1,j-1,k}^\dagger c_{i,j,k} + 2t_0\sigma_z c_{i,j,k}^\dagger c_{i,j,k} + \textnormal{h.c.}\Big).
\end{split}
\end{align}
The two-band Hamiltonian considered in our work contains quadratic terms of the form $H\sim (k_x+ik_y)^2\sigma_++h.c.$ That is, these terms yield the overlap between wave functions whose angular momentum quantum number must differ by 2. For low-lying energies, this would correspond to the s and d orbitals. When strain is applied, the hopping terms are modified as shown by the general formula in Eq. (\ref{ts}) of the main text. Here, we will expand on the explicit calculation. 

Consider the term containing hopping in the $x$ direction: $-c_{i+1,j,k}^\dagger c_{i,j,k}(t_1\sigma_x+t_0\sigma_z)$. In this direction, the change in length due to strain is given by $\delta r_x=a(u_{xx}dx+u_{xy}dy+u_{xz}dz)$, the lattice vector component is $a_x=a\partial_x$ and $a$ is the length of the lattice vector. We will set the derivative $\frac{\partial t}{\partial r}=\gamma\frac{t}{a}$ where $\gamma$ is the Gr\"{u}neisen parameter. Putting these components together and using the dot-product rule $\langle\partial_i, dx^j\rangle=\delta_i^j$, we get
\begin{align*}
    -(t_1\sigma_x+t_0\sigma_z)\rightarrow -(t_1(1+\gamma u_{xx})\sigma_x+t_0(1+\gamma u_{xx})\sigma_z).
\end{align*}
A similar analysis yields the $y$  and $z$ components
\begin{align*}
    (t_1\sigma_x-t_0\sigma_z)&\rightarrow (t_1(1+\gamma u_{yy})\sigma_x-t_0(1+\gamma u_{yy})\sigma_z),\\
    t_3\sigma_z &\rightarrow t_3(1+\gamma u_{zz})\sigma_z.
\end{align*}
For the next-nearest neighbor hopping term along the $k_x+k_y$ direction, we have ${\bf \delta r}=\frac{a}{2\sqrt{2}}(u_{xx}+u_{yy}+2u_{xy})({\bf dx}+{\bf dy})$ and ${\bf a}=\frac{a}{\sqrt{2}}({\bf \partial_x}+{\bf \partial_y})$. Similarly, in the $k_x-k_y$ direction, ${\bf \delta r}=\frac{a}{2\sqrt{2}}(u_{xx}+u_{yy}-2u_{xy})({\bf dx}-{\bf dy})$ and ${\bf a}=\frac{a}{\sqrt{2}}({\bf \partial_x}-{\bf \partial_y})$. This yields the modified terms
\begin{align*}
    t_2\sigma_y &\rightarrow t_2\big(1+\frac{\gamma}{2}(u_{xx}+u_{yy}+2u_{xy})\big)\sigma_y, \\ - t_2\sigma_y &\rightarrow - t_2\big(1+\frac{\gamma}{2}(u_{xx}+u_{yy}-2u_{xy})\big)\sigma_y. 
\end{align*}
On transforming back to $k$ space with the modified hopping parameters, and expanding the resultant Hamiltonian around $(0,0,\pm\pi/2)$ yields the low-lying Hamiltonian below
\begin{align}
\begin{split}
    H&=t_1\big(k_x^2-k_y^2-2\gamma(u_{xx}-u_{yy})\big)\sigma_x-2t_2(2k_xk_y-2\gamma_{xy})\sigma_y\\
    &+2t_3\big(k_z\pm \frac{\pi}{2}\pm\gamma u_{zz}\frac{\pi}{2}\big)\sigma_z+2\gamma t_0(u_{xx}+u_{yy})\sigma_z.
\end{split}
\end{align}
A slightly modified version of this Hamiltonian is used in Eq. (\ref{sth}).

We have notably not used the second part of Eq. (\ref{ts}) which would include deformations out of the plane in the $z$ direction and hence take into account first order derivatives in the $z$ direction. However, these terms are forbidden by symmetry. As mentioned before, the hopping terms are overlaps of orbitals that differ in angular momentum by 2. Such orbitals are even in the $z$-coordinate, thus rendering the expectation values of first order derivatives to vanish. 
\bibliography{references}
\end{document}